\documentstyle[twocolumn,prl,aps,amsfonts,amssymb,epsfig,floats]{revtex}
\begin{document}
\title{Bridging k- and q- Space in the Cuprates: Comparing ARPES and STM
Results}
\author{R.S. Markiewicz}
\address{Physics Department, Northeastern U., Boston MA 02115}
\maketitle
\begin{abstract}
A critical comparison is made between the ARPES-derived spectral function and 
STM studies of Friedel-like oscillations in Bi$_2$Sr$_2$CaCu$_2$O$_{8+\delta}$ 
(Bi2212).  The data can be made approximately consistent, provided that
(a) the elastic scattering seen in ARPES is predominantly small-angle
scattering and (b) the `peak' feature seen in ARPES is really a dispersive
`bright spot', smeared into a line by limited energy resolution; these are
the `bright spots' which control the quasiparticle interferences.
However, there is no indication of bilayer splitting in the STM data.  
\end{abstract}
%****
\narrowtext
%****

Scanning tunneling microscopy (STM) studies\cite{Howl,Hoff2} in Bi$_2$Sr$_2
$CaCu$_2$O$_{8+\delta}$ (Bi2212) find striking periodic 
patterns in real-space local density of states (dos) maps.  A number of
periodicities are found, with different orientations and dispersing with
binding energy $\omega$ below the Fermi level\cite{Hoff2,KMc}.  While a
number of models have been proposed for this effect\cite{PDDH,PSV,Han}, 
here some consequences of a particular quasiparticle interference (QPI)
model\cite{DH,Hoff2} will be explored.  The QPI model suggests that the
periodicities arise from quasiparticle interference effects, similar to the 
Friedel-like oscillations observed\cite{Crom} on clean metal surfaces,
near a step edge or point impurity.  This model has been extremely
successful in predicting an array of periodicities and their
dispersions\cite{Hoff2,KMc}, and moreover extracting both a Fermi surface
$E(\vec k)$and a superconducting gap $\Delta (\vec k)$ which are in good
agreement with those found by angle-resolved photoemission spectroscopy
(ARPES).  

The main issue addressed in the present paper is: are ARPES
results consistent with the QPI model? More specifically, whereas
normal-state Friedel oscillations are dominated by `lines' -- nesting of
flat segments of Fermi surface -- the superconducting QPI is dominated by
`points' -- highly localized peaks in the local density of states, herein
called `bright spots'.  While these bright spots provide a detailed
explanation for the STM observations, they should be directly observable
in the ARPES spectra, and they have not so far been reported.  Here it is
shown that this could be a result of finite ARPES resolution blurring the
bright spots into the quasiparticle peak seen in ARPES in the
superconducting state as extending around much of the Fermi surface.

\begin{figure}
   \epsfxsize=0.33\textwidth\epsfbox{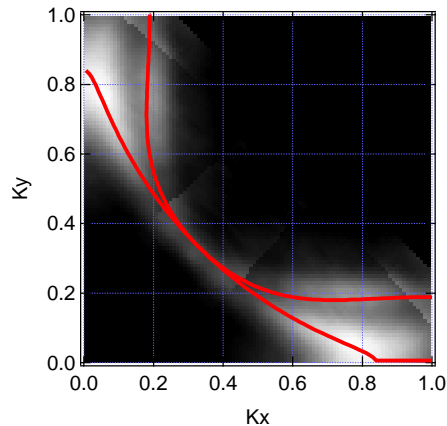}
\vskip0.5cm \caption{Experimental ARPES spectral weight at the Fermi level for 
the sample in the normal state at 100K. Lines show tight 
binding fit.}
\label{fig:0}
\end{figure}

Figure~\ref{fig:0} shows an experimental Fermi surface (FS) map of
an overdoped Pb-doped Bi2212 sample with $T_c=70K$ with strongly
suppressed superstructure\cite{PBog}, taken in the superconducting state. 
The figure shows no trace of the isolated bright spots postulated in the
QPI model (compare, e.g., Fig.~\ref{fig:1} below).  While the STM data
are actually taken on more underdoped samples, the overdoped sample was
chosen as having much sharper spectra (including well resolved bilayer
splitting\cite{Feng}), which should make the bright spots easier to see.
While the ARPES spectrum is multiplied by a matrix element\cite{matri},
there is no reason why this should obscure the bright spots.  

For modelling purposes, the electronic dispersion is fit,
Fig.~\ref{fig:0}, to the form
\begin{eqnarray}
\epsilon_k=-2t(c_x+c_y)-4t'c_xc_y-4t''(c_x^2+c_y^2-1)
\nonumber \\
\pm {t_z\over 2}({c_x-c_y\over 2})^2,
\label{eq:1}
\end{eqnarray}
with $c_i=\cos{k_ia}$, and parameters
%[Bob:] $t=0.326eV$, $t'/t=-0.276$,
%$t''=0.03eV$, $t_z=0.2eV$ ($n$=2), and chemical potential $\mu =-0.38eV$; 
%[Pasha:] 
%$t=0.3eV$, $t'=-0.11eV$, $t''=0.0332eV$, $t_z=0.225eV$ ($n$=1), and 
%chemical potential $\mu =-0.4602eV$.  
%[Pasha:] (x-0.325/ for tz=0 curve)
$t=0.3eV$, $t'=-0.11eV$, $t''=0.028eV$, $t_z=0.24eV$, and chemical potential 
$\mu =-0.44eV$.  While the fit assumes
one hole-like and one electron-like FS, the experimental spectra are so
broad that comparable fits could be made assuming two hole Fermi surfaces.  

\begin{figure*}[t!]
   \epsfxsize=0.88\textwidth\epsfbox{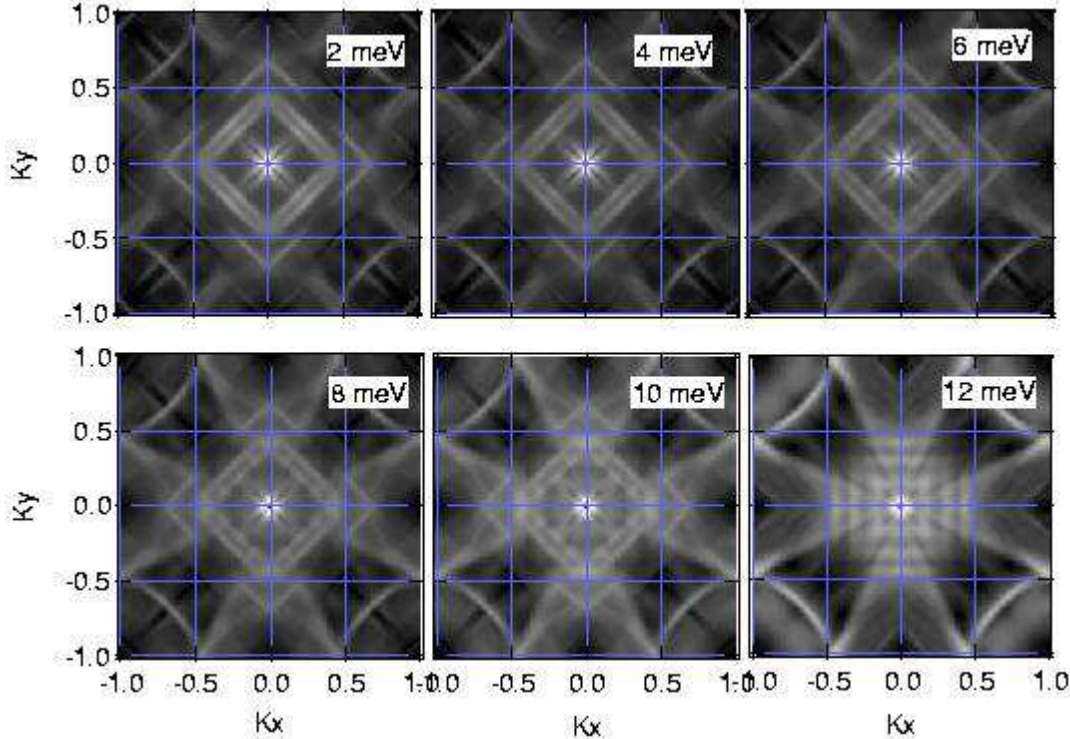}
\vskip0.5cm
\caption{Q-Maps constructed directly from convolution of measured ARPES spectral
functions for a series of frequencies $\omega$ from 2 to 12 meV.}
\label{fig:4ak}
\end{figure*}

A more direct comparison with STM results can be made.  Indeed, the Fourier 
transform of these STM oscillations (which is here called a `q-map') is
approximately\cite{DH} given by the convolution of the ARPES spectral
function at wave number $\vec k$ with that at $\vec k+\vec q$, averaged
over $\vec k$ (see Eq.~\ref{eq:3a1}, below).  
Figure~\ref{fig:4ak} presents an {\it experimental} reconstruction 
of this quantity, derived from convolution of ARPES data at 
a series of energies $\omega$ = 2-12meV, below the superconducting gap $\Delta_
0$.  These data were taken from the same sample as in Fig.~\ref{fig:0};
the analysis was kindly carried out by P. Bogdanov\cite{PBo}.
In the direct convolution, these features are superposed upon a large, 
featureless peak near $(\pi ,\pi )$; to enhance contrast, a derivative of the 
convolution spectrum is displayed.  While there is considerable variation of 
relative intensities of the features with energy, the spectra are characterized 
by patterns of extended lines which do not shift significantly with $\omega$.
This is just the pattern expected from the convolution of the broadened ARPES
spectra illustrated in Fig.~\ref{fig:0}.  Since the ARPES spectral function is 
an extended arc, the convolution gives rise to a q-map consisting of
similar arcs.  

The ARPES spectra should show enhanced scattering at the special points
found by Wang and Lee\cite{DH}, due to a high local density of states.  
To understand the absence of these sharp peaks responsible for the STM
features, it is necessary to calculate model theoretical spectra. 
Since STM measures a local density of states, the resulting q-map is
itself a perturbative correction to the spectral function, which can be
divided into a potential source term $V(q)$ and an
electronic response.  The source term is assumed to be some unknown function
of the impurity distribution, NPS, etc., while the electronic response can
be analyzed for the `spectrum' of strongly coupled wave numbers.  The
response to a point impurity, considered as a superposition of magnetic
and potential scatterers, has been studied in detail for d-wave
superconductors\cite{BFlat}.  A simple expression is possible when the
scattering is weak (Born limit):
\begin{eqnarray}
Im(\delta G(\vec q,\omega ))=\sum_{\vec k}Im(V_{\vec q}G(\vec k+\vec q,\omega )
G(\vec k,\omega ))
\nonumber \\
=V''_{\vec q}\bar\chi'(\vec q,\omega )+V'_{\vec q}\bar\chi''(\vec
q,\omega),
\label{eq:2a}
\end{eqnarray}
where $V_{\vec q}=V'_{\vec q}+iV''_{\vec q}$, and similarly for $\bar\chi
(\vec q,\omega)$ with response function
\begin{equation}
\bar\chi(q,\omega )=\sum_kG(k,\omega)G(k+q,\omega ). 
\label{eq:3d}
\end{equation}
For comparison to ARPES, it is convenient to decompose $\bar\chi'=
\bar\chi'_1+\bar\chi'_2$, with
\begin{eqnarray}
\bar\chi'_1(\vec q,\omega )=\sum_{\vec k}Re(G(\vec k+\vec q,\omega
))Re(G(\vec k, \omega ))
\label{eq:3a1}
\end{eqnarray}
\begin{eqnarray} 
\bar\chi'_2(\vec q,\omega )=
-\sum_{\vec k}Im(G(\vec k+\vec q,\omega ))Im(G(\vec k,\omega )),
\label{eq:3a2}
\end{eqnarray}
The significance of this separation comes from the fact that the term
$\bar\chi'_2$ can be expressed as a convolution of the ARPES 
spectral weight $A=-Im(G)/\pi$ with itself, and hence can be numerically 
reconstructed directly from the ARPES spectrum.  

To account for the large broadening found in the ARPES spectra, the
Green's function is calculated in an Eliashberg approach, with allowance
for elastic scattering.  To minimize pairbreaking effects\cite{MSV}, it is
assumed that the elastic scattering is predominantly small-angle 
scattering ($V_{\vec q}=V_1\delta (\hat q)$), which is not pairbreaking.
(Strong small angle scattering has been postulated in a number of
studies of the cuprates\cite{AAA,Piet,KuZ,GCast,MD,AVar}.)  
In this case, the superconducting state Green's function
becomes\cite{Maki,MSV}
\begin{equation}
G(\vec k,\omega )={\omega Z_{\vec k}+\xi_{\vec k}\over (\omega^2-\Delta_{\vec k}
^2)Z_{\vec k}^2-\xi_{\vec k}^2}
\label{eq:2e}
\end{equation}
with $\xi_{\vec k}=\epsilon_{\vec k}-\epsilon_F$, 
\begin{equation}
Z_{\vec 
k}=1+{\Sigma_{1,\vec k}\over\sqrt{\Delta_{\vec k}^2-\omega^2}},
\label{eq:2dab}
\end{equation}
$\Sigma_{1,\vec k}=n_IN_{\vec k}(0)|V_1|^2$,
and a d-wave gap is assumed $\Delta_{\vec k}=\Delta_0
(c_x-c_y)/2$.  In the overdoped Bi-2212 sample, the measured gap
$\Delta_0(0)$ is 15 meV.
Fits to the superconducting state dispersion give
$\Sigma_1\simeq 20meV$ (the angle dependence of $\Sigma$ is
neglected). As $Im\Sigma$ appears to increase as
doping is decreased, a value $\Sigma_1=30meV$ is assumed for
comparison with STM results.

From Eq.~\ref{eq:2e}, $G(\vec k,\omega )$ develops an imaginary part when
$\omega >\Delta_{\vec k}$.  Near this threshold, $Z_{\vec k}\sim
i\Sigma^{el}_I/\sqrt{\omega^2-\Delta_{\vec k}^2}$, and 
\begin{equation}
A(\vec k,\omega )=
                {\omega\Sigma^{el}_I/\pi\over (\xi_{\vec k}^2+\Sigma^{el2}_I)
\sqrt{\omega^2-\Delta_{\vec k}^2}}
\label{eq:2f}
\end{equation}
if $\omega\ge\Delta_{\vec k}$.  Thus, the spectral weight actually {\it
diverges at the 'bright spots'}, the special points which satisfy both
$\xi_{\vec k}=0$ and $\tilde\omega =\tilde\Delta_k$.  (This is different
from the clean limit ($Z_{\vec k}=1$), where the spectral weight is
proportional to the local density of states, which peaks at the bright
spots.)

Figure~\ref{fig:1} shows the corresponding constant energy maps of $A$ at
several binding energies, including a small pairbreaking\cite{MKSTM},
$\Sigma_2=0.5meV$, plus giving $\omega$ a small imaginary part,
$\delta_{\omega}=1meV$.  The bright spots are clearly visible, while the
accompanying arcs approximately superpose on the bare dispersion at all
energies.

\begin{figure}
   \epsfxsize=0.33\textwidth\epsfbox{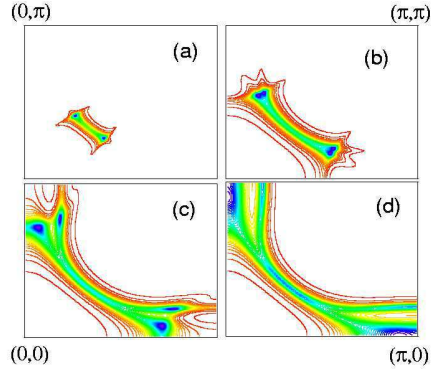}
\vskip0.5cm 
\caption{Model ARPES spectral functions, at $\omega /\Delta$ = 0.2 (a), 0.5 (b),
0.8 (c), and 1.0 (d).}
\label{fig:1}
\end{figure}

\begin{figure}
   \epsfxsize=0.33\textwidth\epsfbox{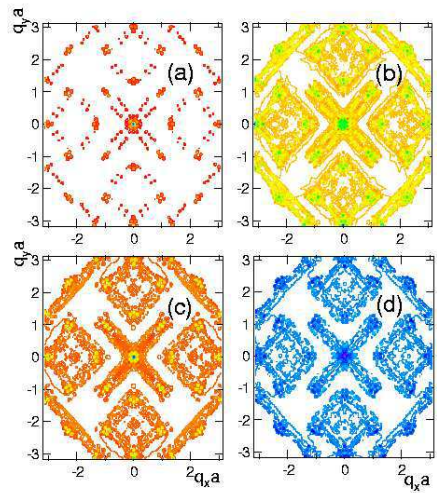}
\vskip0.5cm
\caption{Q-Maps for $\omega$ = 0.5 $\Delta$, $\Sigma_1=0.003eV$:
$-\bar\chi'_2$ (a), $-\bar\chi'$ (b), $-\bar\chi'+\bar\chi''$ (c), and
$-\bar\chi'-\bar\chi''$ (d).}
\label{fig:9}
\end{figure}

Once the Green's function is known, the STM susceptibility,
Eq.~\ref{eq:3d}, can be calculated.  Figure~\ref{fig:9} compares several
different contributions to the q-map, comparing $\bar\chi'_2$ (a) with the
full $\bar\chi'$ (b), and then combining $\bar\chi'$ with $\bar\chi''$.
To estimate the importance of this contribution, it is assumed for
illustrative purposes that $V'_q=\pm V''_q$ in Fig.~\ref{fig:9}c,d.  It
can be seen that the convolution term is dominated by the bright spots,
and reproduces the patterns found in STM measurements.  
Inclusion of the additional terms modifies the details of the spectral
function convolution, enhancing features from the nesting of FS arcs while
making the bright spots stand out somewhat less from the background, but
does not lead to any significant shifts of spectral features.

\begin{figure}
   \epsfxsize=0.33\textwidth\epsfbox{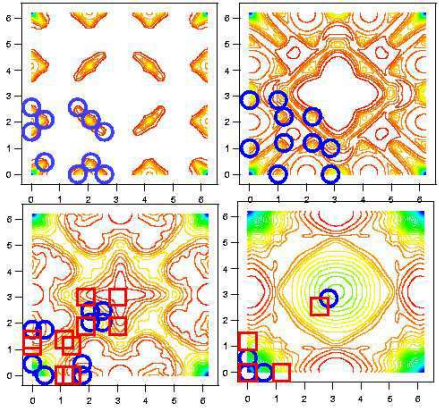}
\vskip0.5cm
\caption{Series of calculated STM q-maps corresponding to pairbreaking model
of Fig.~\protect\ref{fig:1}, for $\omega /\Delta$ = (a) 0.2, (b) 0.5, (c) 0.8,
and (d) 1.0.  Circles (squares) = calculated positions of convolution
intensity corresponding to the most intense (second most intense) peaks in
ARPES intensity.  Note that in this and subsequent q-maps, the origin has 
been shifted with respect to Figs.~\protect\ref{fig:4ak}
and~\protect\ref{fig:8}.} 
\label{fig:6a}
\end{figure}

Figure~\ref{fig:6a} shows a series of q-maps at various excitation
energies $\omega$, corresponding to the constant energy maps of
Fig.~\ref{fig:1}.  For simplicity, only the $\bar\chi'_2$ contribution is
included.  The general form of the resulting q-maps can readily be
understood as a convolution of the bright spot peaks, exactly as in the
experimental analysis\cite{KMc}.  The calculated positions of the bright spot
convolutions are shown by the circles and squares in Fig.~\ref{fig:6a}; at
the higher energies, where the bilayer splitting is evident, the circles
represent the stronger peaks associated with the antibonding band (nearer
the VHS), the squares the weaker bonding band peaks.  (Additional 
scattering from bonding to antibonding bright spots is automatically
included in the calculated maps, but not illustrated by symbols.)  
While the resulting spectra include weak extended arcs, 
which shift little with $\omega$, the bright spot features are quite
prominent and strongly $\omega$-dependent, yielding a q-map in good 
agreement with Ref.~\onlinecite{Hoff2}.  
For $\omega >0.5\Delta$, the agreement with experiment is less good.  The
spots from the bonding and antibonding bands overlap, leading to smoother
spectra without clearly resolved bright spots.  In contrast, the STM
studies {\it do not find any features associated with the antibonding
bands}, so the bonding band bright spots persist to higher energies.

\begin{figure}
   \epsfxsize=0.33\textwidth\epsfbox{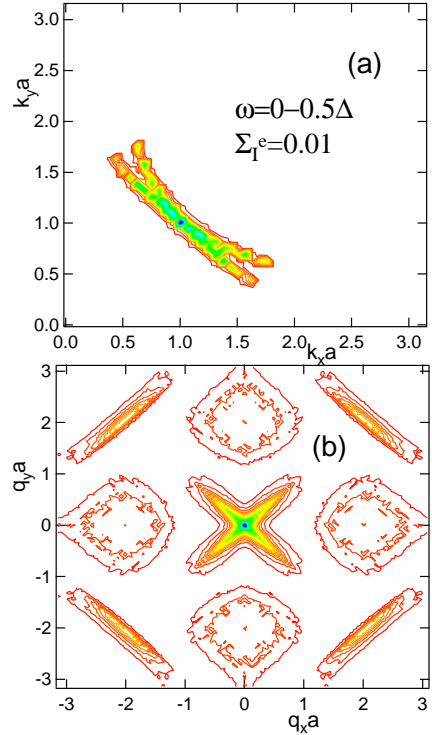}
\vskip0.5cm
\caption{Calculated ARPES (a) and STM q-maps (b), assuming limited energy
resolution $\omega =0-0.5\Delta$. }
\label{fig:8}
\end{figure}

While the experimental ARPES spectral functions show sharp quasiparticle
peaks in the superconducting state, the bright spots are not clearly
resolved.  This can be understood as due to the finite energy resolution.
Thus, Fig.~\ref{fig:8}a shows te result of averaging the MD spectra over
the range $0\le\omega\le 0.5\Delta_0$.  The resulting strips of roughly
constant intensity are in much better agreement with the measured ARPES
spectra.  Note that the experimental resolution in Fig.~\ref{fig:0} is
$\sim 15 meV\sim\Delta_0$.  The resulting convolution, Fig.~\ref{fig:8}b,
is in good agreement with the experimental convolutions of
Fig.~\ref{fig:4ak}.

There is one striking discrepancy between the model calculations and the
STM maps: the ARPES spectra find a bilayer splitting while STM does not.
In particular, model q-maps find both bilayer split bands, with the most
intense feature coming from the (antibonding) band nearest the Van Hove
singularity.  On the other hand, the STM derived q-maps were
inverted\cite{KMc} to reconstruct the Fermi surface, and {\it only a
single Fermi surface section was found}, corresponding to the bonding
band. While the bilayer splitting is most easily observed in overdoped and Pb
substituted samples, it is found\cite{bil,DanD2} that in underdoped
samples the bilayer splitting does not decrease,
although the spectra broaden. 

In conclusion, a detailed comparison of recent ARPES and STM measurements
in BISCO has been provided.  It is found that the most striking features
of the STM results can be understood in terms of `bright spot'
quasiparticles, but that conclusive evidence will have to await higher
resolution ARPES measurements.  In the clean limit, the ARPES spectral
weight of the bright spots has been estimated.  In particular, it is
predicted that the ARPES `peak' feature is really an image of these
dispersive bright spots, smeared into an extended streak by finite energy
resolution.
Residual discrepancies remain between the STM-derived and ARPES Fermi surfaces, 
in particular the absence of bilayer splitting in the STM results.

Acknowledgments:
This work was begun while I was on sabbatical at Stanford.  I thank
Z.X. Shen, P. Bogdanov, A. Lanzara, J.C. Davis, A. Kapitulnik, and M.
Greven for many stimulating conversations, and Z.X. Shen and P. Bogdanov
for permission to use Figs.~\ref{fig:1},~\ref{fig:4ak}.

\end{document}